\documentclass[submission,copyright,creativecommons]{eptcs}
\providecommand{\event}{SOS 2007} 
\usepackage{breakurl} 

\usepackage{graphics} 
\usepackage{graphicx} \usepackage{subfigure}
\usepackage{epsfig} 
\usepackage{amsmath} 
\usepackage{amssymb} 

\usepackage{t1enc}

\newtheorem{definition}{Definition}

\title{Stochastic Analysis of Synchronization in a Supermarket Refrigeration System}
\author{John Leth\thanks{This work is supported by the Southern Denmark Growth Forum and the European Regional Development Fund under the project ``Smart \& Cool''.} 
\institute{Department of Electronic Systems, Aalborg University, Fredrik Bajers Vej 7C, 9220 Aalborg, Denmark} 
\email{jjl@es.aau.dk}
\and
Rafael Wisniewski
\institute{Department of Electronic Systems, Aalborg University, Fredrik Bajers Vej 7C, 9220 Aalborg, Denmark}
\email{raf@es.aau.dk}
\and
Jakob Rasmussen
\institute{Department of Mathematical Sciences, Aalborg University, Fredrik Bajers Vej 7G, 9220 Aalborg, Denmark}
\email{jgr@math.aau.dk}
\and
Henrik Schioler
\institute{Department of Electronic Systems, Aalborg University, Fredrik Bajers Vej 7C, 9220 Aalborg, Denmark}
\email{henrik@es.aau.dk}
}
\def\titlerunning{Stochastic Analysis of Synchronization in a Supermarket Refrigeration System}
\def\authorrunning{J. Leth, R. Wisniewski, J. Rasmussen, H. Schioler}
\begin{document}

\maketitle

\begin{abstract}      
Display cases in supermarket systems often exhibit synchronization, in which the
expansion valves in the display cases turn on and off at exactly the
same time. The study of the influence of switching noise on synchronization in supermarket refrigeration systems
is the subject matter of this work.  For this purpose, we model it as
a hybrid system, for which synchronization corresponds to a periodic
trajectory.
Subsequently, we
investigate the influence of switching noise. We develop a statistical
method for computing an intensity function, which measures how often
the refrigeration system stays synchronized. By analyzing the
intensity, we conclude that the increase in measurement uncertainty
yields the decrease at the prevalence of synchronization. 
\end{abstract}

\section{Introduction}



In a supermarket refrigeration, the interaction between temperature controllers leads to a
synchronization of the display cases in which the expansion
valves in the display cases turn on at the same time. This phenomenon causes high wear of compressors \cite{LTWI:07}.


In this article, we apply the concepts in \cite{lethCDC:12} to study a system in \cite{raf:14}.
More precisely, we combine the hybrid system model of a refrigeration system presented in \cite{raf:14} with the proposed method for modeling switching noise presented in \cite{lethCDC:12}, and show that this formalism is able to characterize de-synchronizing behavior as a consequence of inaccuracy of temperature measurement.    
The affect of this inaccuracy has been utilized in a patent that proposes to adjust the cut-in and cut-out temperatures for the refrigeration entities to de-synchronize them \cite{PatentDanfoss}.

 

The definitions of
synchronization have been formulated in
\cite{Blekhman_Fradkov_Nijmeijer_Pogromsky_1997}, and numerous examples of synchronization have been
analyzed in \cite{Pikovsky_Maistrenko_2003}. In \cite{raf:14}, the synchronization of a supermarket refrigeration
system was studied as periodic trajectories of a hybrid system with a state
space consisting of a disjoint union of polyhedral sets and discrete transitions realized by reset maps defined on the facets
of the polyhedral sets.

However in real life applications, the switching instances are not deterministic, they are
influenced by system and measurement noise.  Therefore, we suggest a
practical method rooted in stochastic analysis \cite{lethCDC:12}, where noise is
introduced into the system. For the stochastic analysis, we
``thicken'' the switching surfaces to a neighborhood around them and
formulate a probability measure of the switching in such a way that
the longer the trajectory stays within the neighborhood, the higher is
the probability of switching. As a result, this method provides the
means of computing intensity plots. By analyzing them, we conclude
that contrarily to the deterministic behavior, the system synchronizes
intermittently. In addition, we infer that the prevalence of
synchronization increases with decreasing uncertainties of the
temperature measurements. 
In conclusion, the method of intensity plots provides a tangible
test of whether and how frequently synchronization occurs in a
refrigeration system.

The article is organized as follows. In Section~\ref{Sec:RefSys} and Section~\ref{Sec:SupermarketSwitchedSys} we recall, from \cite{raf:14}, a simple model of a refrigeration system and demonstrate that this model coincide with the notion of a hybrid system, technical details are given in the appendix. Section~\ref{sec:saos} start by introducing the notion of switching uncertainity from \cite{lethCDC:12}, which then is used to incorporate inaccuracy of temperature measurement into the model. Subsequently a synchronization analysis of the refrigeration system is performed.


\section{Refrigeration system}
\label{Sec:RefSys}
\begin{figure}[hpbt]
  \centering
   \includegraphics[width=0.4\textwidth]{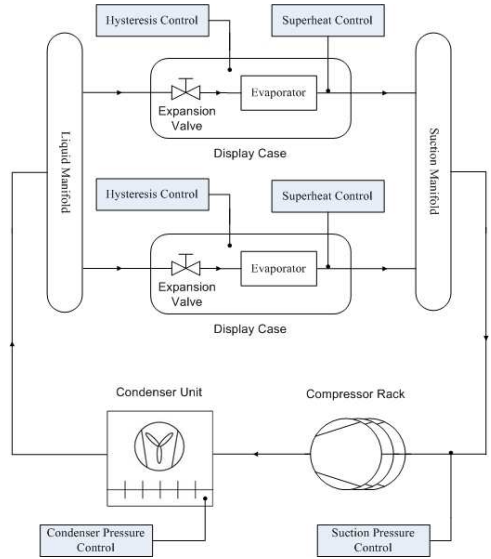}
   \caption{Layout of a simple supermarket refrigeration system.}
   \label{fig:syst_layout}
 \end{figure}

First, a brief description of a refrigeration cycle of a supermarket refrigeration
systems with display cases and compressors connected in parallel, see figure~\ref{fig:syst_layout} for a graphic layout.

The compressors, which maintain the
flow of refrigerant, compress refrigerant drained from the suction manifold. Subsequently,
the refrigerant passes through the condenser and flows into the liquid
manifold. Each display case is equipped with an expansion valve,
through which the refrigerant flows into the evaporator in the display
case. In the evaporator, the refrigerant absorbs heat from the
foodstuffs. As a result, it changes its phase from liquid to
gas. Finally, the vaporized refrigerant flows back into the suction
manifold. 


In a typical supermarket refrigeration system, the temperature in each
display case is controlled by a hysteresis controller that opens the
expansion valve when the air temperature $T$ (measured near to the
foodstuffs) reaches a predefined upper temperature limit $T^u$. The
valve stays open until $T$ decreases to the lower temperature limit
$T^l$. At this point, the controller closes the valve again.
Practice reveals that if the display cases are similar,
the hysteresis controllers have tendency to synchronize the display
cases \cite{LTWI:07}. It means that the air temperatures $T_i$ for $i
\in \{1, \hdots, N\}$, where $N$ is the number of display cases, tend to
match as time progresses.

In the sequel we discuss, for simplicity, a model of a refrigeration system that consists of only two identical display cases and a compressor. The dynamics of the air temperature $T_i$ for display case $i\in\{1,2\}$ and the suction
pressure $P$ for the system of two display cases are governed
by the following system of equations, 

\begin{equation}
  \dot{x}=\xi_\delta(x)=A_\delta x+B_\delta,
  \label{Eq:SimplifiedRefrigerationModel}
\end{equation}
where
\[
A_\delta=
\begin{bmatrix}
  -a-\delta_1c&0&\delta_1d\\
  0&-a-\delta_2c&\delta_2d\\
  0&0&-\alpha
\end{bmatrix},~
B_\delta=
\begin{bmatrix}
  b+e\delta_1\\
  b+e\delta_2\\
  \beta+\delta_1+\delta_2
\end{bmatrix},~ x=
\begin{bmatrix}
  T_1\\
  T_2\\
  P
\end{bmatrix},
\]
with $a,b,c,d,e,\alpha,\beta$ constants whose specific values
are provided in Appendix \ref{AppendixA}, and $\delta=(\delta_1,\delta_2) \in \mathbf 2^2$ with $\mathbf 2
= \{0,1\}$ and $\delta_i \in \{0,1\}$ the switching parameter for display
case $i$; it indicates whether the expansion valve is closed ($\delta_i=0$) or open ($\delta_i=1$). The switching law is given by
the hysteresis control:
\begin{equation}
{ \delta_i=
  \begin{cases}
    1 &\text{if}  ~T_i\geq T_{i}^u\\
    0 &\text{if}  ~T_i\leq T_i^l\\
    \delta_i &\text{if}~T_i^l< T_i<T_i^u,
  \end{cases} 
}
\label{eq:hysterisisAdhock}
\end{equation}
where $T_i^u$ and $T_i^l$ are respectively the predefined upper and lower
temperature limits for display case $i$. By convention, $\delta_i=0$ for any initial condition $T_i(t_0)=T_i^0\in]T_i^l,T_i^u[$. Such an initial condition is assumed 
 throughout  this paper; hence,~\eqref{eq:hysterisisAdhock} is well defined.

\section{Refrgeration system as a hybrid system}
\label{Sec:SupermarketSwitchedSys}

\begin{figure}[hpbt]
  \centering
  \includegraphics[width=0.3\textwidth]{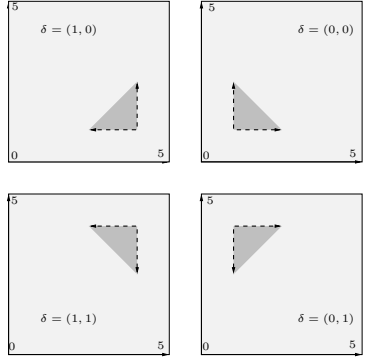}
  \caption{The state space of the refrigeration system consisting of
    two display cases is illustrated. Here, the pressure axis is  suppressed, and
    $(T^l_i,T^u_i)=(0,5)$. The direction of the vector field
    $\xi_\delta$ is indicated by the dark shaded triangles (see Proposition~1 in \cite{raf:14}).}\label{fig:delta}
\end{figure}

Consider the following scenario. Let $x(t_0)~\in~]T_1^l,T_1^u[\times]T_2^l,T_2^u[\times \mathbb
R_+$, and $\delta = (0,0)$; thereby, both
display cases are initially switched off. 
Suppose that at time $t$, the air temperature $T_i$ of the $i$th
display case reaches the upper temperature limit $T_i^u$, then the
$i$th display case is switched on, and $\delta_i = 1$. This scenario
indicates that the refrigeration system \eqref{Eq:SimplifiedRefrigerationModel} comprises four dynamical
systems, each defined on a copy of the polyhedral set
\begin{align}\label{eq:q}
Q = [T_1^l,T_1^u]\times[T_2^l,T_2^u]\times \mathbb{R}_+,
\end{align}
as illustrated in Fig.~\ref{fig:delta}.
A discrete transition between these four systems takes place whenever
a trajectory reaches one of the following four facets of $Q$
\begin{align*} F^{\alpha}_1 &= F^{\alpha}_1 (Q) = \delta^{\alpha}
  [T_1^l,T_1^u]\times[T_2^l,T_2^u]\times \mathbb R_+ ~\hbox{ and }~ \\
  F^{\alpha}_2 & = F^{\alpha}_2(Q) = [T_1^l,T_1^u]\times
  \delta^{\alpha} [T_2^l,T_2^u]\times \mathbb R_+, \end{align*}
with $\alpha \in \mathbf 2$ and $\delta^0 [a,b] = \{a\}$, $\delta^1
[a,b] = \{b\}$.

Moreover, the transitions between subsystems can be describe by eight reset maps $R_i(\delta): F_i^{l_i(i,\delta)} \times
\{\delta\} \to F_i^{l_i(i,\delta)} \times \{l(i,\delta)\}$ defined
by $R_i(\delta)\left(x,\delta\right) = (x,l(i,\delta))$ with
\[
l(i,\delta) = (l_1(i,\delta),l_2(i,\delta)) =
\left \{
\begin{matrix} (\delta_1+1,\delta_2) & \hbox{if} & i=1 \\
    (\delta_1,\delta_2+1) & \hbox{if} & i=2,
  \end{matrix}
\right.
\]
where the results of the summation are computed modulo
$2$. Intuitively, the map $l$ takes a polyhedral set enumerated by
$\delta$ to the future polyhedral set. The variable $i$ indicates that
the discrete transition takes place when the temperature $T_i$ reaches its upper or
lower boundary. The domain of a reset map, will be referred to as a switching surface. Specifically, a switching surface is a
facet of one of the polyhedral sets making up the state space of the
refrigeration system~\eqref{Eq:SimplifiedRefrigerationModel}.


The above construction allows us to identify the refrigeration system as a switched hybrid system. More precisely, let ${\cal P}=\{P_{\delta}~|~P_{\delta} = Q \times\{\delta\},~\delta \in {\mathbf 2}^2\}$ consist of the four polyhedral sets making up the state space, ${\cal S}=\{\xi_{\delta}~|~\xi_\delta(x)=A_\delta x+B_\delta,~\delta \in {\mathbf 2}^2\}$ consist of the four dynamical systems, and $\mathcal R = \{R_i(\delta)~|~(i,\delta) \in \{1, 2\} \times \mathbf 2^2\}$ consist of the eight reset maps. Then the triple $({\cal P}, {\cal S}, {\cal R})$ constitute a hybrid system as defined in Appendix~\ref{sec:ss}. Hence the behavior of the system is expressed by means of a (hybrid) trajectory which heuristically can be described simply as the union of trajectories generated by four local dynamical systems, see Appendix~\ref{app:tra} for a precise describtion. 

With this notion at hand we say that a refrigeration system exhibit asymptotic synchronization if there exists a
$(T,l)$-periodic trajectory which is asymptotically stable in $X^*$ 
\cite[Definition 13.3]{Had:06}, where $X^*$ denote the quotient space $X/\sim$ with $X = \bigcup_{\delta
  \in {\mathbf 2}^2} P_{\delta}$ and $\sim \subset X \times X$ the
equivalence relation, \cite{bredon}, generated by the reset maps in $\mathcal R$. For a detailed explanation see \cite{raf:14}, where it is also shown that the refrigeration system generates an asymptotically stable $(T_p,2)$-periodic trajectory, with $T_p\approx 261$, lying on the diagonal of $P_{(0,0)}$ and $P_{(1,1)}$.  

\begin{figure}[hpbt]
  \centering
  \includegraphics[width=0.4\textwidth]{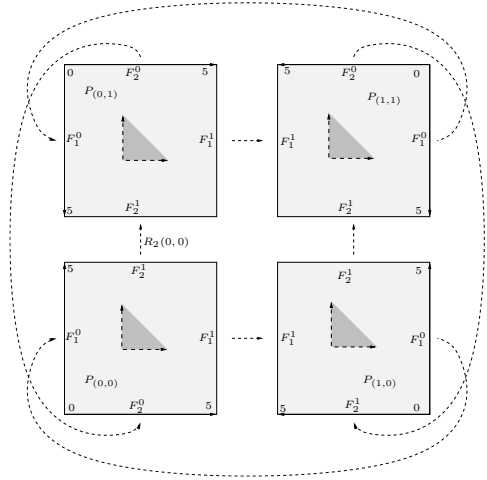}
  \caption{The $T_1T_2$-state space of the refrigeration system
    consists of two display cases. The reset maps are indicated by
    the stippled lines (see Fig.~\ref{fig:delta} and its caption for
    further explanation). The pressure axis has been suppressed; thus,
    each $P_{\delta} = Q \times \{\delta\}$ is illustrated by a
    square. By abuse of notation, the facets of $P_{\delta}$ are
    denoted by $F_i^{\alpha}$ (instead of $F_i^{\alpha} \times
    \{\delta\}$).}\label{fig:SynchronisationCube}
\end{figure}

The (hybrid) refrigeration system with two display cases is
illustrated in Fig.~\ref{fig:SynchronisationCube}. Here, each element
of $\cal P$ has been (orthogonally) projected onto the
$(T_1,T_2)$-space. Hence, the polyhedral sets $P_{\delta}$ are
represented by cubes. The three cubes $P_{(0,1)}$, $P_{(1,0)}$,
$P_{(1,1)}$ have been vertically and/or horizontally reflected
(compare with Fig.~\ref{fig:delta}). The stippled lines in the
drawing indicate the reset maps in $\cal R$.

\section{Stochastic analysis of synchronisation}\label{sec:saos}
So far, the discrete transition from a local system to another has been described
as deterministic. In other words, it take place with probability one
when a trajectory ``hits'' a switching surface. However,
from a practical point of view, this causes a problem.  For
instance, any sampling step will result in the trajectory
``hitting'' the switching surface with probability zero. On account of
model uncertainty, noise,  and most of all the inaccuracy of temperature measurement, the
switching surface should therefore be ``thickened'' by replacing each switching
surface with an open neighborhood of it.

In the sequel, we incorporate the ``thickening of switching surfaces''
into our model of the refrigeration system. Specifically, we
construct an open neighborhood around the switching surfaces within which
a probability measure on each trajectory is proposed. Afterwards, we
use this measure to describe the probability of a discrete transition, in such a
way that the longer a trajectory stays within the neighborhood the
higher becomes the probability of a transition. This makes it possible to
devise a method for analyzing the typical behavior of the system, as
well as a way of visualizing it.

Let $U$ be a stochastic variable uniformly distributed on
$[-\epsilon,\epsilon]$ with density function $
p(u)=\mathbf{I}_{[-\epsilon,\epsilon]}(u)/2\epsilon$, where
$\mathbf{I}_A$ denotes the indicator function of a set $A$. Consequently, the
distribution of $U$ can be described by the survivor function 
\[
S(u) =
1-\int_{-\epsilon}^u p(v) dv= \begin{cases} 
1 & u < -\epsilon \\
1-(u+\epsilon)/2\epsilon & |u| \leq \epsilon \\
0 & u > \epsilon,
\end{cases}
\] 
which is the
probability of $U\geq u$, $\textbf{P}(U\geq u)$. The distribution of
$U$ can also be described by its conditional intensity (or hazard)
function
$h(u)=p(u)/S(u)=\mathbf{I}_{[-\epsilon,\epsilon]}(u)/(\epsilon-u)$.
Heuristically, a small variation of $h(u)$ is the probability of $U$ being in a small region around $u$
conditional on $U$ not being smaller than $u$, $\textbf{P}(U\in du~|~
U\geq u)$.
The conditional intensity function, for short intensity, turns out to
be a convenient starting point for defining stochastic transitions
between the different local dynamical systems.

To this end, we derive an intensity function suitable for our purpose. Recall
the notation introduced in Section~\ref{sec:ss}, and let
$F(\delta)=F_1^{l_1(1,\delta)}\cup F_2^{l_2(2,\delta)}$ denote the
union of switching surfaces in the polyhedral set
$P_\delta$. Furthermore, let $F(\delta)^\epsilon$ denote the
$\epsilon$-neighborhood of $F(\delta)$ in $\mathbb{R}^3$.

For $x\in F(\delta)^\epsilon$, let $F_i^{l_i(i,\delta)}$ be a switching
surface within $\epsilon$ (Hausdorff) distance to $x$, and define
$x_{(i,\delta)}=x-\pi_{(i,\delta)}(x)$, where
$\pi_{(i,\delta)}$ is the orthogonal projection onto
$F_i^{l_i(i,\delta)}$. Hence, when $x_{(i,\delta)}$ is non-zero, it is a normal
vector to $F_i^{l_i(i,\delta)}$ (more precisely, to the affine hull of
$F_i^{l_i(i,\delta)}$). It points into $P_\delta$ when
$x\in P_{\delta}-F_i^{l_i(i,\delta)}$ and out of $P_\delta$ when
$x\in F(\delta)^\epsilon-P_{\delta}$.
Let $n_{(i,\delta)}$ denote a normal vector to $F_i^{l_i(i,\delta)}$
which points out of $P_\delta$. By means of the above
quantities, we define the parameter
$
u_{(i,\delta)}(x)=|x_{(i,\delta)}|\text{sign}\langle n_{(i,\delta)}, x_{(i,\delta)} \rangle
$
where $x_{(i,\delta)}=x_{(i,\delta)}(x)$ is regarded as a function of
$x$. 

Now let $\gamma(t;k)$ denote a trajectory (see Appendix~\ref{app:tra}) of the refrigeration system, and assume that $\gamma(t;k)$ follows system $\delta$, i.e.,
$\dot{\gamma}(t;k)=\xi_\delta(\gamma(t;k))$. The intensity function
$h_{(i,\delta)}$ for switching from system $\delta$ to system
$l(i,\delta)$ at $\gamma(t;k)$ can now be defined as
\begin{align*}
h_{(i,\delta)}(\gamma(t;k))=
\begin{cases}
0
& u_{ij}(\gamma(t;k))<-\epsilon\\
h(u_{ij}(\gamma(t;k)))
& |u_{ij}(\gamma(t;k))|\leq \epsilon\\
\infty
& u_{ij}(\gamma(t;k))>\epsilon.
\end{cases}
\end{align*}
Hence, $h_{(i,\delta)}(\gamma(t;k))$ is the intensity of the signed
orthogonal distance from $\gamma(t;k)$ to the switching surface
$F_i^{l_i(i,\delta)}$ in the $\epsilon$-neighborhood of $P_\delta$. As
a result, any trajectory whose intersection with the
$\epsilon$-neighborhood of $F_i^{l_i(i,\delta)}$ is contained in a
normal subspace to $F_i^{l_i(i,\delta)}$ will switch according to the
uniform distribution on the restriction of the trajectory to the
$\epsilon$-neighborhood.

The above construction successfully copes with switching for which a trajectory
reaches one switching region at a time. To deal with the situation
where a point of switching is within distance $\epsilon$ to both
switching surfaces, we will assume that switching to each of the
systems will happen independently. With this assumption, we
immediately conclude that if $\gamma(t;k)$ follows system
$\delta$, and is within distance $\epsilon$ from both
$F_2^{l_2(2,\delta)}$ and $F_2^{l_2(2,\delta)}$ then the intensity
function $h_{\delta}$ for switching from system $\delta$ to system
$l(i,\delta)$ at $\gamma(t;k)$ is given by $
h_{\delta}(\gamma(t;k)(t))=h_{(1,\delta)}(\gamma(t;k))+h_{(2,\delta)}(\gamma(t;k))$. Thus, if  the discrete transition occurs at $\gamma(t;k)$, the switch to the system
$l(i,\delta)$ happens with probability $
h_{(i,\delta)}(\gamma(t;k))/h_{(\delta)}(\gamma(t;k)).  $
Equivalently, we can allow switching to happen according to all of the
intensities $h_{(1,\delta)}(\gamma(t;k))$ and
$h_{(2,\delta)}(\gamma(t;k))$ independently of each other and
disregard all the switchings except the first one which determines the
switching.


Based on the above theory and numerical simulations, we now conduct synchronization analysis of the refrigeration system when
switching noise is present. To begin with, we describe how the theory
is implemented in simulation. For this, we follow Algorithm 7.4III in
\cite[p.\ 260]{daley}.

When a trajectory enters the switching region $F(\delta)^{\epsilon}$,
an exponentially distributed variable with mean one is
simulated. Thereafter, for each sample, the intensity is calculated,
and subsequently the integral of the intensity is computed
inductively. Afterwards, the exponentially distributed variable is
compared with the number obtained by the integral computation
\cite[p.\ 258 (Lemma 7.4II)]{daley}. 
 As a consequence, the switching occurs at the sample
point where the integral exceed the exponentially distributed
variable. 

In the above setup, the system trajectories become stochastic; hence,
we can apply concepts such as mean values to describe the system
behavior. For a trajectory $(\gamma,\mathcal{T}_\infty)$, $A\subset X$
and $I=[t',t'']\subset[0,\infty)$, let $Z_\gamma(A,I)$ be the arc
length from $t'$ to $t''$ of $\gamma$ inside $A$. As a consequence,
$\gamma\mapsto Z_\gamma(A,I)$ defines a non-negative random variable,
whose mean $\mathbb{E}[Z(A,I)]$ describes the curve intensity relative
to $(A,I)$. The curve intensity measures the typical behavior of
trajectories by the mean length of trajectories on $I$ inside $A$.

We illustrate the typical behavior of the refrigeration system with
two display cases by simulation of the curve intensity. For this
purpose, we use the following algorithm: (1) Fix a time interval
$I\subset [0,\infty)$ and a sufficiently regular subset $A\subset X$. (2)
Simulate $n$ trajectories $\gamma_1,\dots\gamma_n$. (3) Divide $A$
into small subsets $A_i$. (4) Calculate
$\mathbb{E}[Z(A_i,I)]\approx\frac{1}{n}\sum_jZ_{\gamma_j}(A_i,I)$
where $Z_{\gamma_j}(A_i,I)$ is approximated by the number of sampling
points (on trajectories) falling in $A_i$.

\begin{figure}[hpbt]
  \centering
  \subfigure{\includegraphics[width=0.24\textwidth]{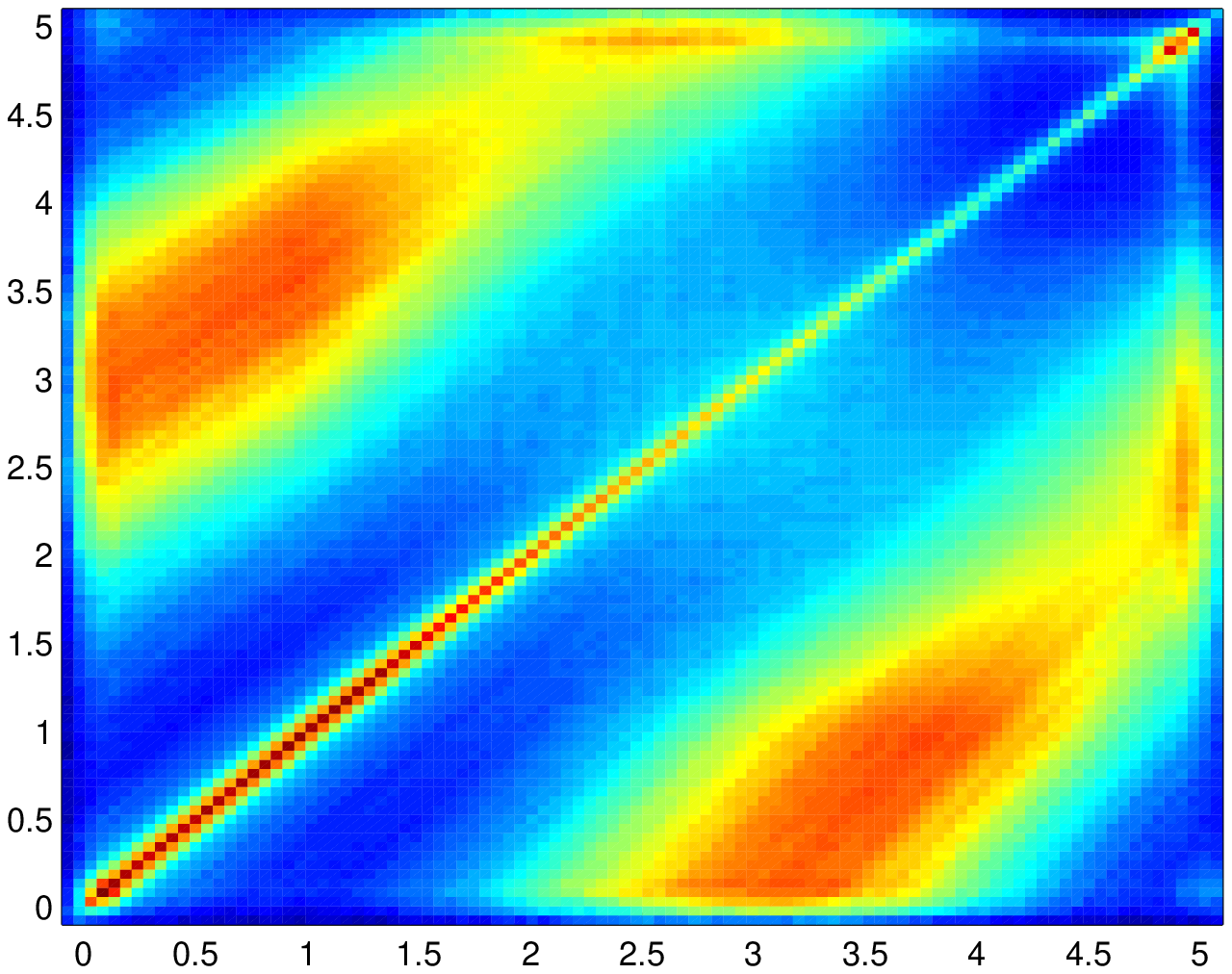}}
  \subfigure{\includegraphics[width=0.24\textwidth]{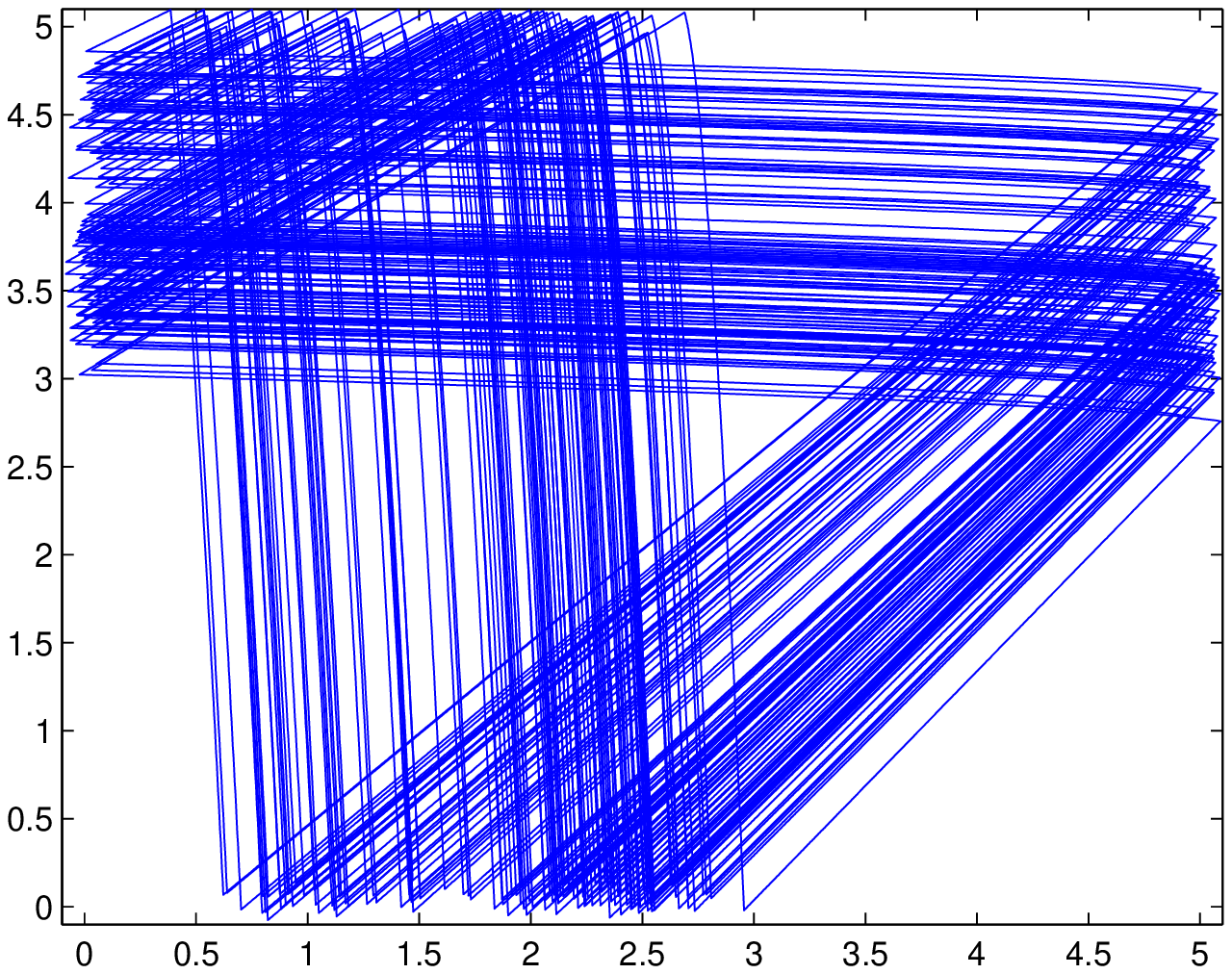}}
  \subfigure{\includegraphics[width=0.24\textwidth]{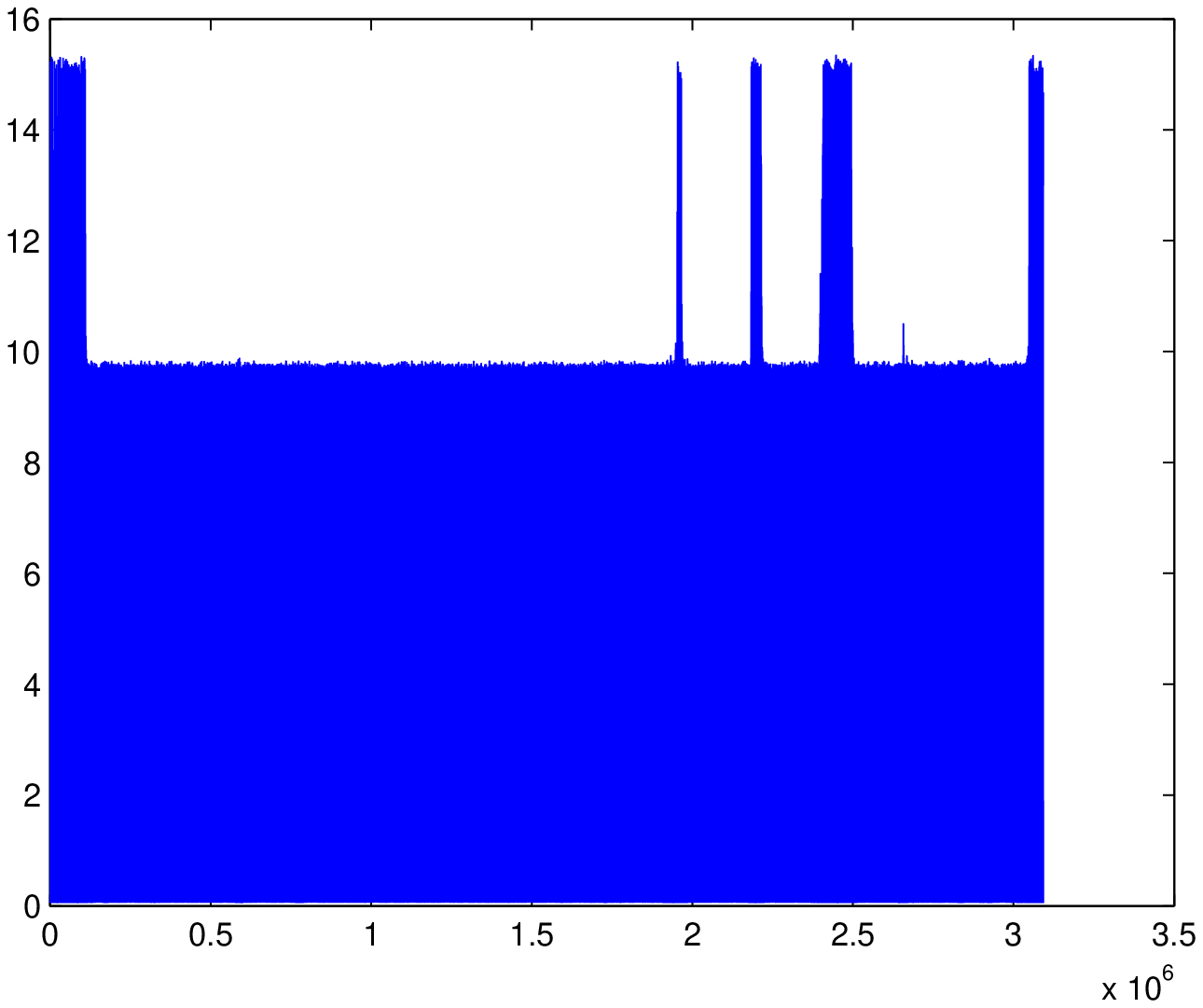}}
  \subfigure{\includegraphics[width=0.24\textwidth]{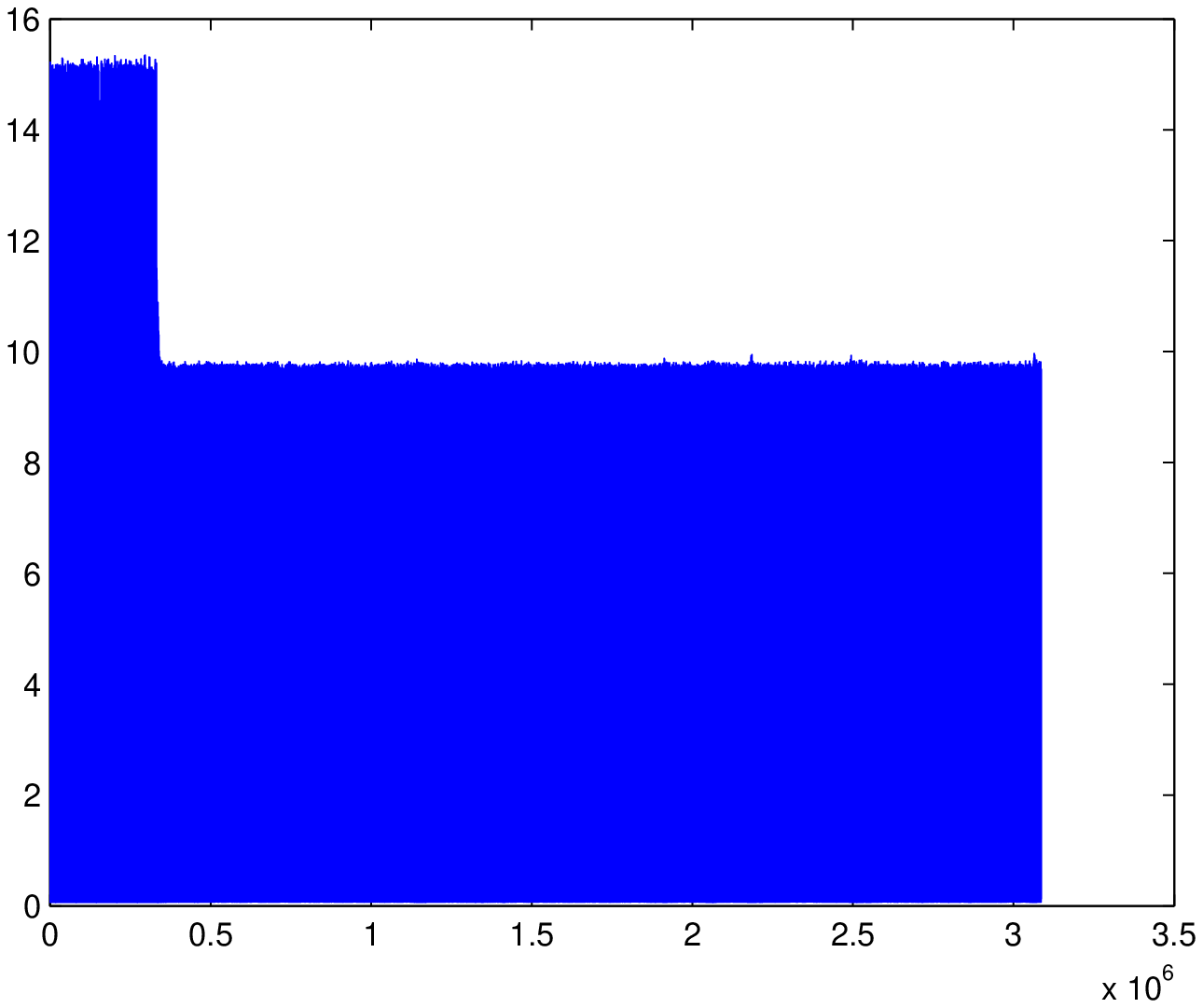}}
  \subfigure{\includegraphics[width=0.24\textwidth]{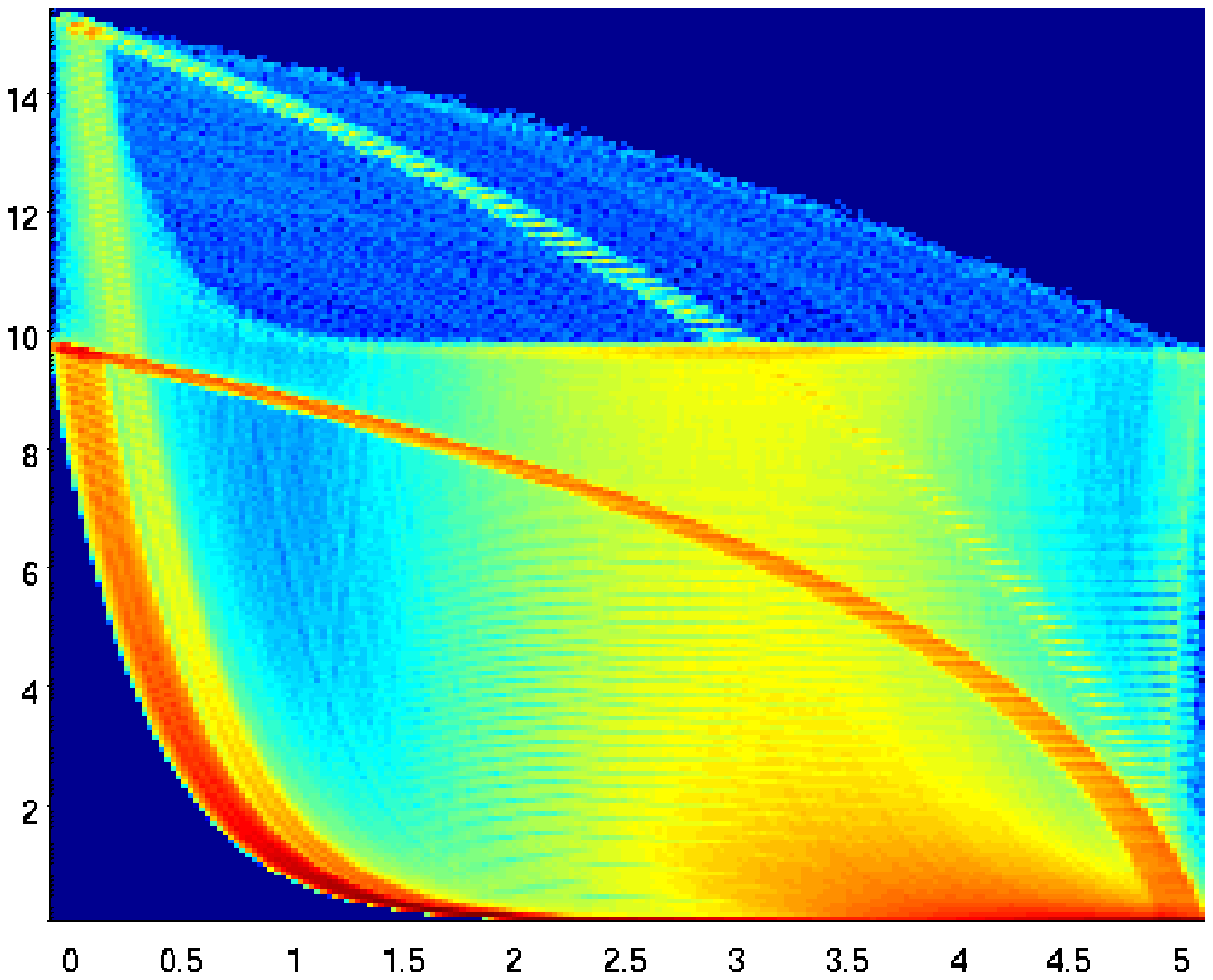}}
  \subfigure{\includegraphics[width=0.24\textwidth]{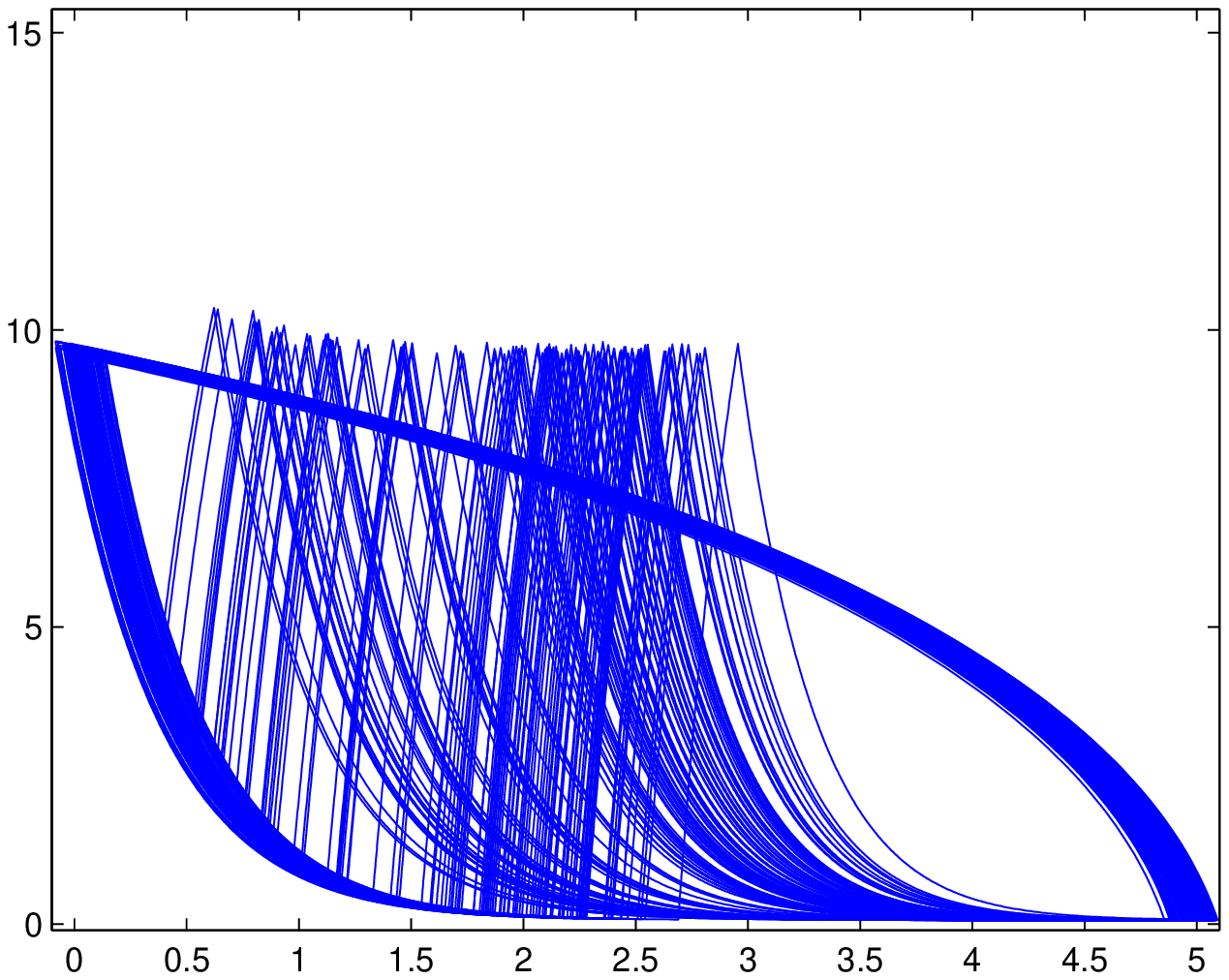}}
  \subfigure{\includegraphics[width=0.24\textwidth]{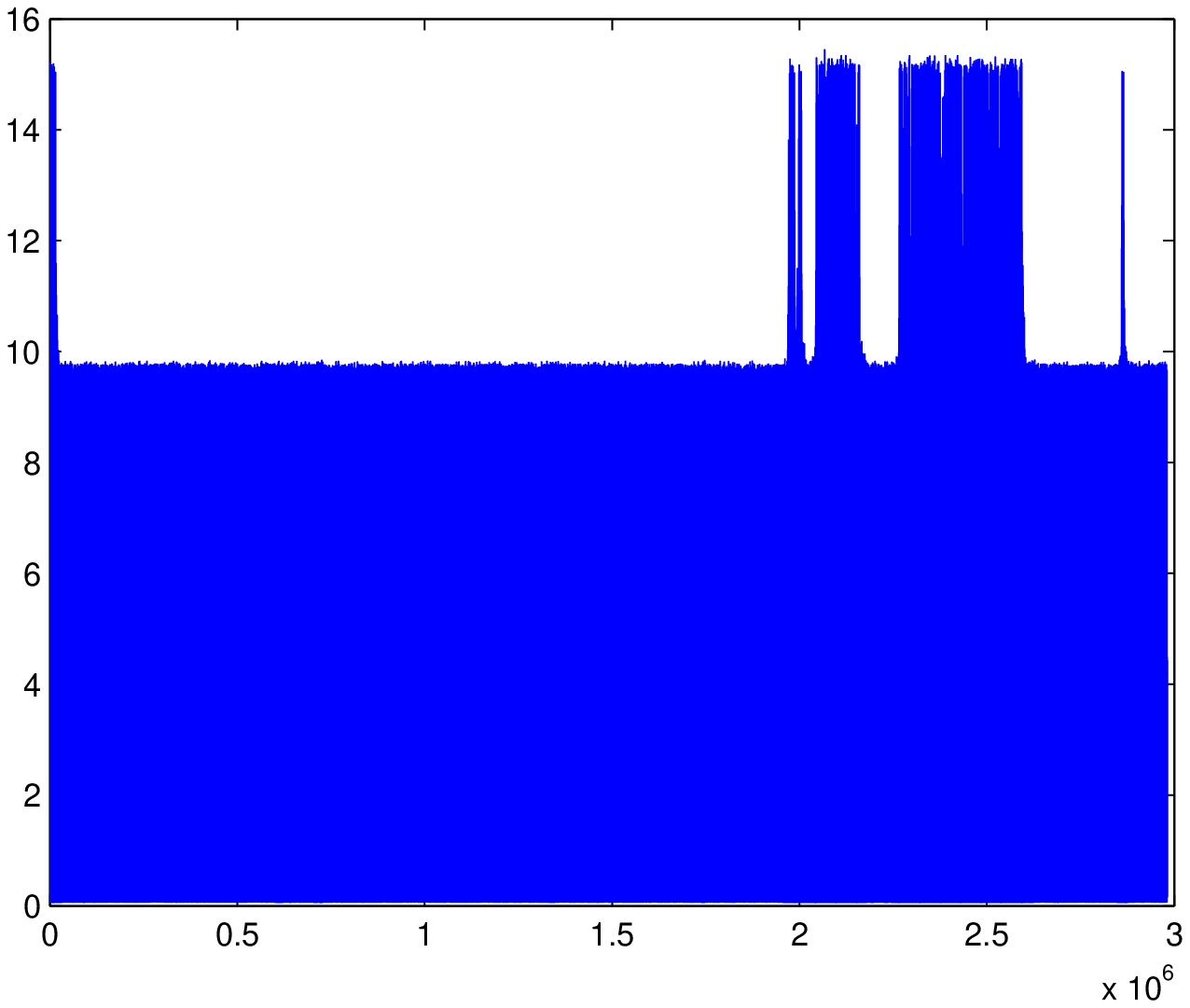}}
  \subfigure{\includegraphics[width=0.24\textwidth]{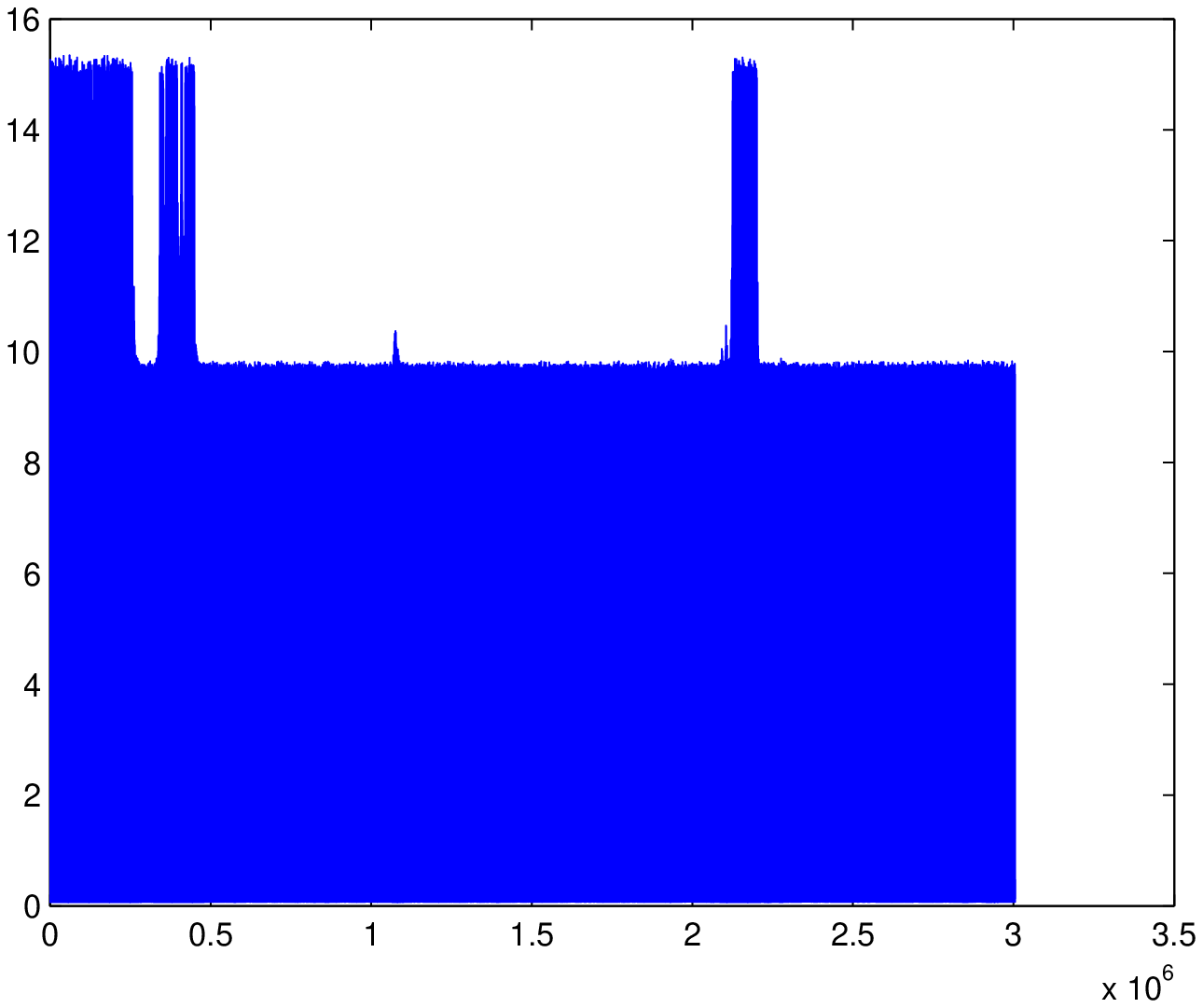}}
  \caption{ Each of the plots in column one and two are illustrated on various projections of the space obtained by identifying the four copies of $Q$ in \eqref{eq:q} used to define the state space of the refrigeration system.  
	The first column represents intensity plots,
    $\epsilon=0.1,~I=[0,10^6]$, based on $n=10$ trajectories
     at the point $(0,0,15.23)$ on the asymptotically stable $(T_p,2)$-periodic trajectory found in \cite{raf:14}
		(top: $T_1T_2$-plane, bottom: $T_1P$-plane). The
    second column represents one of the ten trajectories in a
    subinterval of $I$ (top: $T_1T_2$-plane, bottom: $T_1P$-plane) shown in the time interval $[10^6,1.1\cdot 10^6]$. The last two
    columns represent four of the ten simulations and show the sample-time
    evolution of suction pressure $P$. 
  }\label{fig:plot}
\end{figure}
Intensity plots for $I=[0,10^6]$, $\epsilon=0.1$ and $n = 10$ have
been generated for initial values both within and outside the basin of
attraction of the asymptotically stable $(T_p,2)$-periodic trajectory found in \cite{raf:14}. The resulting intensity plots are similar to the one
illustrated in the first column of Figure~\ref{fig:plot}.
Contrarily to the deterministic behavior, the system synchronizes
occasionally over the time interval $I$, as seen in the last two columns
of Figure~\ref{fig:plot}, where the pressure peaks at $15.23$ correspond to the presence
of the $(T_p,2)$-periodic trajectory (synchronization).
The behavior of the system in the remaining time is
indicated in the first column of Figure~\ref{fig:plot} as the colored
area except for the diagonal (top) and yellow/green circle through
$(0,15.23)$, and in the last two columns as pressure peaks at $10$. We
remark that the peaks at $10$ do not correspond to another limit
cycle, as shown in the second column of Figure~\ref{fig:plot} (in fact
this correspond to the interval $[10^6,1.1~10^6]$ of the plot in
the right lower corner).

Further simulations reveal that the time spent in the
$(T_p,2)$-periodic trajectory is highly dependent on $\epsilon$. For
$\epsilon=0.1$, relatively little time is spent in the
$(T_p,2)$-periodic trajectory, whereas for $\epsilon=0.01$, the
behavior closely matches that of the deterministic case. More
precisely, for $\epsilon=0.01$ and initial values in the basin of
attraction, the system rapidly converges to the $(T_p,2)$-periodic
trajectory and stays close to it with a high probability, while for
initial values outside the basin of attraction, the system has a small
probability of converging to the $(T_p,2)$-periodic trajectory within
the time interval $I$. As a consequence, the relative frequency of
appearance of the $(T_p,2)$- trajectory can be described as a
monotonically decreasing function $f(\epsilon) \in [0, 1]$, where
$f(0) = 1$ corresponds to the deterministic case.

In summary, uncertainties of the temperature measurements influence
synchronization and can be used to design de-synchronization
scheme. We conclude that adding noise as above to
temperature measurements de-synchronizes the refrigeration system:
An increase in measurement uncertainty yields a decrease of the prevalence of
synchronization.



\section{Conclusion}
In this paper, we have investigate the influence of switching noise on the synchronization phenomenon in
supermarket refrigeration systems. 
We have developed a
numerical method for computing intensity plots. By analyzing them, we
have concluded that the time the refrigeration system stays
synchronized is dependent on uncertainties of the temperature
measurements. A greater measurement uncertainty yields a smaller
accumulated time in which the refrigeration system is in
synchronization.

\section{Appendix}

\subsection{Model of refrigeration system}
\label{AppendixA}
The mathematical model presented here is a summary of the model
developed in \cite{WL1:08}. For the $i$th display case, dynamics of the air temperature
$T_{air,i}$ can be formulated as
\begin{subequations}
\begin{align}
  &\frac{{d}T_{air,i}}{{d}t}=\frac{\dot{Q}_{goods-air,i} +
    \dot{Q}_{load,i}-\delta_i\dot{Q}_{e,max,i}}{\left(1+\frac{UA_{goods-air,i}}{UA_{air-wall,i}}\right) 
    M_{wall,i} C_{p,wall,i}} ~~\hbox{ with } \label{eq:TairApp} \\
  &T_{wall,i} =
  T_{air,i}-\frac{\dot{Q}_{goods-air,i}+\dot{Q}_{load,i}}{UA_{air-wall,i}}, \\
  &\dot{Q}_{goods-air,i} = UA_{goods-air,i}(T_{g0,i}-T_{air,i}), \\
  &\dot{Q}_{e,max,i} =
  UA_{wall-ref,max,i}(T_{wall,i}-a_{T}P_{suc}-b_{T}), 
\end{align} 
\end{subequations}
where the process parameters are specified in Table~\ref{tab:para}, and $\delta_i \in \{0, 1\}$ is the switch parameter for
the $i$th display case. When $\delta_i = 0$, the $i$th expansion valve
is switched off, whereas when $\delta_i = 1$ it is switched on. The suction manifold dynamics is governed by the differential equation
\begin{align}\label{eq:PsucApp}
  \frac{{d}P_{suc}}{{d}t} = \frac{\textstyle\sum_{i = 1}^k \delta_i
    \dot{m}_{0,i}+\dot{m}_{r,const}-\dot{V}_{comp} (a_{\rho}
    P_{suc}+b_{\rho})}{V_{suc}\cdot \nabla\rho_{suc0}},
\end{align}
where $N$ is the number of display cases, and $\dot
m_{0,i} = \dot m_0$ for $i \in \{1, \hdots, N\}$. 

We denote $T_i = T_{air,i}$ and $P = P_{suc}$ and write
the dynamics of the air temperature and suction pressure 
in the concise form (with the process constants in \eqref{eq:TairApp}
collected in $a,b,c,d,e,\alpha,\beta$ and then replaced by their numerical values)
\begin{subequations}\label{eq:RefrigerationApp}
\begin{align}
  \dot{T}_i&=-aT_i+b-\delta_i(cT_i-dP-e) \label{eq:dott} \\
  &=-0.0019T_i+0.0244-\delta_i(-0.0012T_i+0.0506P+0.1065),\nonumber \\
  \dot{P}&=-\alpha P+\beta+\delta_1+\delta_2
  =-0.056P+0.0038+\delta_1+\delta_2.
\end{align}
\end{subequations}

\begin{table}[hpbt]
  \begin{center}
{\tiny
    \begin{tabular}{|l||l||l||l|}\hline
      \multicolumn{4}{|l|} {\textbf{Display cases}}\\
      \hline $UA_{wall-ref,max}$  &   500  $\frac{J}{s\cdot K}$ & $T_{g0}$
      &
      3.0  $^0C$ \\
      \hline $UA_{goods-air}$ & 300  $\frac{J}{s\cdot
        K}$ & $\dot m_0$ & 1.0  $kg/s$ \\
      \hline $UA_{air-wall}$ & 500 $\frac{J}{s\cdot K}$ & $\dot Q_{load}$ & 3000 $J/s$ \\
      \hline
      $\dot m_{r,const}$ & 0.2  $\frac{kg}{s}$ & $M_{wall}$ & 260  $kg$ \\
      \hline $\nabla\rho_{suc0}$ & 4.6  $\frac{kg}{m^3bar}$ & $C_{p,wall}$ & 385  $\frac{J}{kg \cdot K}$\\
      \hline \multicolumn{4}{|l|} {\textit{The same parameters has been
          used for all
          display cases.}}\\
      \hline \multicolumn{4}{|l|} {\textbf{Compressor}}\\\hline
      $\dot V_{comp}$ & 0.28  $\frac{m^3}{s}$ & \multicolumn{2}{l|}{} \\
      \hline
      \multicolumn{4}{|l|} {\textbf{Suction manifold}}\\
      \hline
      $V_{suc}$ & 5.00 $m^3$ & \multicolumn{2}{l|}{} \\
      \hline
      \multicolumn{4}{|l|} {\textbf{Air temperature control}}\\
      \hline $T_i^l$ & 0.00  $^0C$ & $T_i^u$ & 5.00  $^0C$ \\
      \hline
      \multicolumn{4}{|l|} {\textbf{Coefficients}}\\
      \hline $a_{T}=-16.2072$ &  $b_T={-41.9095}$ & $a_{\rho}=4.6$ &
      $b_{\rho}=0.4$ \\
      \hline
    \end{tabular}}
    \caption{Parameters for a simplified supermarket refrigeration
      system}\label{tab:para}
  \end{center}
\end{table}

\subsection{Hybrid systems}
\label{sec:ss}
A detailed study of the hybrid system presented below can be found in \cite{let:14}. We write $F \prec P$ if $F$ is a face of the polyhedral set $P$.  A map $f : P \to P'$ is polyhedral if 1) it is a continuous injection, and 2) for any $F \prec P$
there is $F' \prec P'$ with $\dim(F) = \dim(F')$ such that $f(F) \subseteq F'$.
\begin{definition}[Hybrid System]
  For finite index sets $J$ and $D$, a hybrid system (of dimension
  $n$) is a triple $({\cal P}, {\cal S}, {\cal R})=({\cal P}_D, {\cal
    S}, {\cal R}_J)$, where
  \begin{enumerate}
  \item $\mathcal{P}=\{P_{\delta} \subset \mathbb{R}^n~|~P_{\delta}
    \hbox{ a polyhedral set}, \dim(P_{\delta})= n,~\delta \in D\}$ is
    a family of polyhedral sets.
  \item ${\cal S}=\{\xi_{\delta}: P_{\delta} \rightarrow \mathbb{R}^n~|~P_{\delta} \in {\cal P},~{\delta} \in D \}$ is a
    family of smooth vector fields.
  \item ${\cal R}=\{R_{j}: F \rightarrow F'~| ~F \prec P\in {\cal P},~
    F' \prec P'\in {\cal P},~\dim(F) = \dim(F') = n-1, ~j \in J\}$ is
    a family of polyhedral maps, called reset maps.
  \end{enumerate}
\end{definition}
After identifying $D$ with a finite subset of $\mathbb R$, we can rewrite  the hybrid system $({\cal P}_D, {\cal S}, {\cal R}_J)$ as
Fig.
\begin{eqnarray*}
  \frac{\mathrm d}{\mathrm d t}( x,  q) &\in& \bar F( x,  q) = \{(\xi_{q}(x),0)|~ x \in P_q\} ~\hbox{ for } ( x, q) \in \bar C ,\hbox{ and }\\
  (x,q)^+ \in  \{R_k(x,q)|~x &\in&  \bar G( x,  q)  = \mathrm{dom}
  (R_k) \prec P_{q}\} ~\hbox{ for } (x,q) \in \bar D,
 \end{eqnarray*}
where $\bar C = \bigcup_{\delta \in D} (P_{\delta} \times
 \{\delta\})$ and $\bar D =\bigcup_{\{(\delta,j) \in D \times J~|~
   \mathrm{dom} (R_j) \prec P_{\delta}\}}(\mathrm{dom} (R_j) \times
 \{\delta\})$.  This is precisely the hybrid system in
 \cite{Goebel_Sanfelice_Teel_2009}.

\subsection{Trajectories of a refrigeration system}\label{app:tra}
We bring in a concept of a (hybrid) time domain \cite{Goebel_Sanfelice_Teel_2009}. 
Let $k\in \mathbb{N}\cup
\{\infty\}$; a
subset $\mathcal{T}_k \subset \mathbb R_+ \times \mathbb Z_+$ will be
called a time domain if there exists an increasing sequence
$\{t_i\}_{i\in \{0,\hdots, k\}}$ in $\mathbb R_+\cup\{\infty\}$ such
that
{\small
\[
\mathcal{T}_k = \bigcup_{i \in \{1,\hdots, k\}}T_i \times \{i\}\]
where $T_i= [t_{i-1}, t_{i}] \hbox{ if } i \in \{1, \dots, k-1\}, \hbox{ and }
T_k = \left\{\begin{matrix}[t_{k-1}, t_{k}]  \hbox{ if } t_k <
    \infty \\ [t_{k-1},\infty[   \hbox{ if } t_k =
    \infty.  \end{matrix} \right.$}

Note that $T_i= [t_{i-1}, t_{i}]$ for all $i$ if $k=\infty$.  We
say that the time domain is infinite if $k = \infty$ or $t_k =
\infty $. The sequence $\{t_i\}_{i\in \{0,\hdots, k\}}$ corresponding to a time domain will be called a switching sequence.

\begin{definition}[Trajectory]\label{Definition:Trajectory} A trajectory
  of the hybrid system $({\cal P}_D, {\cal S}, {\cal R}_J)$ is a
  pair $({\cal T}_k,\gamma)$ where $k\in \mathbb{N}\cup \{\infty\}$ is
  fixed, and
  \begin{itemize}
  \item $\mathcal{T}_k \subset \mathbb R_+ \times \mathbb Z_+$ is a
    time domain with corresponding switching sequence $\{t_i\}_{i\in
      \{0, \hdots, k\}}$,
  \item $\gamma:\mathcal{T}_k \to X = \bigcup_{\delta \in D}
    P_{\delta} \times \{\delta\}$ is continuous ($X$ has the disjoint union topology)
    and satisfies:
    \begin{enumerate}
    \item\label{jump} For each $i \in \{1, \hdots, k-1\}$, there exist
      $\delta \neq \delta' \in D$ such that $\gamma (t_{i};i) \in
      \text{bd}(P_{\delta})$, and $\gamma
      (t_{i};i+1)\in\text{bd}(P_{\delta'})$.
    \item\label{cp} For each $i \in \{1, \hdots, k\}$, there exists
      $\delta \in D$ such that the Cauchy problem
      $
        \frac{\partial}{\partial t}\gamma(t;i)=\dot{\gamma}(t;i)
        = \xi_{\delta} (\gamma(t;i)),~\gamma (t_{i-1};i)=x_{i-1}
        \in P_{\delta},
      $
      has a solution on $T_i\subset\mathcal{T}_k$.
    \item For each $i \in \{1, \hdots, k-1\}$, there exists $j\in J$
      such that $R_j(\gamma(t_i;i)) = \gamma(t_i;i+1)$.
    \end{enumerate}
  \end{itemize}
  A trajectory at $x$ is a trajectory $({\cal T}_k,\gamma)$ with
  $\gamma(t_0;1)=x$. By abuse of notation $\gamma$ will sometimes be referred to as a trajectory. 
\end{definition}

The next definition formalizes the notion of a periodic trajectory, which
will be used in defining synchronization of the refrigeration system.
\begin{definition}[$(T,l)$-periodic trajectory]
  Let $(T,l) \in \mathbb R_+ \times \mathbb Z_+$. A trajectory $({\cal
    T}_{k},\gamma)$ is $(T,l)$-periodic (or just periodic) if (1)
  ${\cal T}_{k}$ is an infinite time domain, and (2) for any $i \in
  \{1, \hdots, k\}$ and $t \in p({\cal T}_{k})$, where $p: {\cal T}_{k}
  \to [t_0, \infty[$ is the projection $p(t,i) = t$, we have $\gamma(t+T;i+l)
  = \gamma(t;i)$.
\end{definition}

In particular, if $({\cal T}_{k},\gamma)$ is a $(T,l)$-periodic trajectory, and $T$ is
   nonzero then $p: {\cal T}_{k} \to [t_0, \infty[$ is surjective.

\newpage

\nocite{*}
\bibliographystyle{eptcs}
\bibliography{mybibfile}

\end{document}